\documentclass[aps,prd,preprint,floatfix]{revtex4}
\usepackage{epsfig}

\begin{document}

\preprint{\rightline{ANL-HEP-PR-02-004}}

\title{Quenched lattice QCD at finite isospin density and related theories.}

\author{J.~B.~Kogut}
\address{Dept. of Physics, University of Illinois, 1110 West Green Street,
Urbana, IL 61801-3080, USA}
\author{D.~K.~Sinclair}
\address{HEP Division, Argonne National Laboratory, 9700 South Cass Avenue,
Argonne, IL 60439, USA}

\begin{abstract}
We study quenched QCD at finite chemical potential, $\mu_I$, for the third 
component of isospin and quenched two-colour QCD at finite chemical potential,
$\mu$, for quark number. In contrast to the quenched approximation to QCD at
finite quark-number chemical potential, the quenched approximations to these
theories behave similarly to the full theories. The reason is that these
theories have real positive fermion determinants. In both of these theories
there is some critical chemical potential above which the charge coupled to
the chemical potential is spontaneously broken. In each case, the transition
appears to be second order. We study the scaling properties near the critical
point using scaling functions suggested by effective (chiral) Lagrangians and
find evidence for scaling with mean-field critical exponents in each case.
The subtleties associated with observing the critical scaling of these
theories are discussed.
\end{abstract}
\maketitle

\pagestyle{plain}
\parskip 5pt
\parindent 0.5in

\narrowtext

\section{Introduction}

RHIC at Brookhaven and the CERN heavy-ion program give us the possibility of
producing hadronic matter at high temperatures and finite baryon number
density. Cold nuclear matter -- hadronic matter at finite baryon number density
-- exists in neutron stars. In addition, the bulk properties of large nuclei
should be well described as nuclear matter. Because of Coulomb interactions,
nuclear matter is not only at finite baryon-number density but also finite
(negative) isospin($I_3$) density. Finally, it has been suggested that
sufficiently dense nuclear matter will also have finite strangeness density.
It is therefore of interest to study QCD at finite baryon-number, isospin and
strangeness densities both at zero and finite temperature.

Such finite densities are achieved by introducing a chemical potential for the
relevant charge operator. Unfortunately, the introduction of a finite chemical
potential for quark/baryon number leads to a complex fermion determinant which
precludes the use of standard simulation methods based on importance sampling.
Introducing a chemical potential $\mu_I$ for the third component ($I_3$) of
isospin leaves the determinant real and non-negative. Adding the additional
(small) $I_3$ breaking term needed to observe spontaneous isospin breaking on
a finite (lattice) volume, makes this determinant strictly positive and 
simulations possible. Effective Lagrangian analyses of this theory have 
indicated that it should undergo a phase transition at $\mu_I=m_\pi$ to a
state where $I_3$ is spontaneously broken by a charged pion condensate with
an accompanying Goldstone mode \cite{ss} We are currently performing
simulations of this theory \cite{lattice2001,qcdi}. Adding, in addition, a
finite chemical potential ($\mu_s$) for strangeness again makes the fermion
determinant complex. In this case, however, there are related theories with
real positive fermion determinants which mimic the correct physics for small
$\mu_s$ \cite{lattice2001}.

Until we have a satisfactory way of dealing with finite baryon-number density,
it is of interest to study models which exhibit some of the anticipated
properties of QCD at finite quark/baryon-number density but which have real
positive fermion determinants. One such model is 2-colour QCD at finite
quark-number chemical potential ($\mu$). This theory exhibits diquark
condensation for $\mu$ sufficiently large. \cite{hm,hklm,kts,kshm,aadgg,mnn}
Effective Lagrangians and chiral perturbation theory analyses of this theory
display similar behaviour, and give quantitative predictions for the nature
of this transition and the equation of state in the neighbourhood of this
critical point \cite{kstvz,stv}. These predictions have been validated in the
strong coupling limit of 2-colour QCD at finite $\mu$ by Aloisio {it et al.}
\cite{aadgg}. Formation of such condensates has been suggested for QCD at large
enough $\mu$ \cite{b,bl,arw,rssv,as}. There is, of course, one crucial
difference. The condensate for 2-colour QCD is a colour singlet, the symmetry
breaking is realized in the Goldstone mode and this theory exhibits
superfluidity. For true, 3-colour QCD, the condensate is, of necessity,
coloured, the symmetry breaking is realized in the Higgs mode and this theory
exhibits colour superconductivity.

The quenched approximation, i.e. the approximation of setting the fermion
determinant to unity has proved useful for calculating hadron spectra and
matrix elements. At finite temperatures it shows the deconfinement transition
at which chiral symmetry is restored, as does the full theory. However, since
the order of this transition and the equation of state are flavour dependent,
it yields no useful information on these issues. Where it can be used, its
principal advantage is that it reduces the computing requirements by several
orders of magnitude. For QCD at finite quark-number chemical potential, the
quenched approximation was even more appealing, since it avoids the problem of
the complex fermion determinant. Unfortunately, it was soon discovered that
it does not produce the correct physics \cite{barbour,bbdkmsw,dk}.
Whereas it is believed that as $\mu$ is increased, the first phase transition
should occur for $\mu \sim m_N/3$, the quenched theory showed a transition for
$\mu \approx m_\pi/2$. This was realized to indicate that the quenched theory
should be considered as the zero-flavour limit of a theory with an equal
number of quark flavours with quark-number $+1$ and with quark-number $-1$,
rather than of a theory where all the quarks have quark number $+1$. This was
implicit in this early work \cite{barbour,bbdkmsw,dk}, and was made explicit
in terms of random matrix models by Stephanov \cite{stephanov}.

Since QCD at finite isospin chemical potential has equal numbers of quarks with
$I_3=+\frac{1}{2}$ and with $I_3=-\frac{1}{2}$, it thus should be expected to
admit a sensible quenched approximation. What one immediately realizes is that
earlier studies of quenched QCD at finite quark-number chemical potential
\cite{barbour,bbdkmsw,dk,kls} can now be reinterpreted as studies of quenched
QCD at finite isospin chemical potential.

Now, however, we can include an explicit $I_3$ violating interaction which
makes the Dirac operator positive definite (rather than positive semi-definite)
giving better convergence for our inversion algorithm, and allowing us to
measure the $I_3$-breaking pion condensate directly. Similarly, in 2-colour
QCD, since quarks and antiquarks belong to the same representation (the
fundamental) of $SU(2)_{colour}$, it too should have a sensible quenched
approximation. In fact, we see, it is the same property that gives these
theories real positive fermion determinants that allows quenched
approximations. This is not surprising, since all a positive fermion
determinant does is to reweight the contributions. This contrasts with a
determinant where the sign of the real part changes and contributions from
different configurations can (and in the case of QCD at finite $\mu$ must)
cancel.

Once one has determined that each of these theories undergoes a second order
transition to a state characterized by a condensate which spontaneously breaks
the charge coupled to the chemical potential, it is useful to examine the
scaling properties of the order parameter and certain composite operators in
the vicinity of the critical point and to obtain the critical exponents. This
allows one to write down an equation of state for the system. Such equations
of state are important for modeling neutron stars. Of course there one needs
to work at finite baryon-number chemical potential as well. Effective
Lagrangians and chiral perturbation theory through 1-loop suggest that we
should see mean field scaling near this critical point \cite{ss,kstvz,stv}.

We have measured the pion condensates, chiral condensates and isospin densities
as functions of isospin chemical potential ($\mu_I$) and the explicit isospin
breaking parameter $\lambda$ on $12^3 \times 24$ and $16^4$ quenched QCD gauge
configurations at $\beta=6/g^2=5.7$, and on $8^4$ quenched gauge configurations
at $\beta=5.5$. Both show evidence for the expected mean-field scaling. In
addition we have measured the diquark condensate, chiral condensate and
quark-number density on a set of $8^4$ and $12^4$ quenched 2-colour QCD
configurations with $\beta=4/g^2=2.0$. Here mean-field scaling is again
favored.

In section 2 we present our actions and their relevant symmetries. Section 3
describes our simulations and results. Critical scaling analyses are presented
in section 4. Discussions and conclusions are presented in section 5.

\section{Actions and Symmetries}

The staggered fermion part of the action for lattice QCD with degenerate $u$
and $d$ quarks at a finite chemical potential $\mu_I$ for isospin ($I_3$) is
\begin{equation}
S_f=\sum_{sites} \left[\bar{\chi}[D\!\!\!\!/(\tau_3\mu_I)+m]\chi
                   + i\lambda\epsilon\bar{\chi}\tau_2\chi\right]
\end{equation}
where $D\!\!\!\!/(\mu)$ is the standard staggered $D\!\!\!\!/$ with links in
the $+t$ direction multiplied by $e^{\frac{1}{2}\mu}$ and those in the $-t$
direction multiplied by $e^{-\frac{1}{2}\mu}$. The explicit symmetry breaking
term $i\lambda\epsilon\chi^T\tau_2\chi$, the lattice equivalent of
$i\lambda\bar{\psi}\gamma_5\tau_2\psi$, is in a direction in which the
$I_3$ symmetry is expected to break spontaneously for $\mu_I$ sufficiently
large. This term is necessary in order to observe spontaneous symmetry
breaking from a finite lattice. We will be interested in the limit $\lambda
\rightarrow 0$. We will present a detailed discussion of the symmetries of
this theory in a forthcoming paper on our simulations with dynamical quarks.
The Dirac operator
\begin{equation}
{\cal M}=\left[\begin{array}{cc}
               D\!\!\!\!/(\mu_I)+m         &           \lambda\epsilon       \\
               -\lambda\epsilon            &           D\!\!\!\!/(-\mu_I)+m
               \end{array}                                            \right]
\label{eqn:su3dirac}
\end{equation}
has determinant
\begin{equation}
\det{\cal M}=\det[{\cal A}^\dagger{\cal A}+\lambda^2]
\end{equation}
where we have defined 
\begin{equation}
{\cal A} \equiv D\!\!\!\!/(\mu_I)+m.
\end{equation}
We note that this determinant is positive for $\lambda \ne 0$, as promised. 
Observables we measure include the chiral condensate
\begin{equation}
\langle\bar{\psi}\psi\rangle \Leftrightarrow \langle\bar{\chi}\chi\rangle,
\end{equation}
the charged pion condensate
\begin{equation}
i\langle\bar{\psi}\gamma_5\tau_2\psi\rangle  \Leftrightarrow 
i\langle\bar{\chi}\epsilon\tau_2\chi\rangle
\end{equation}
and the isospin density
\begin{equation}
j_0^3 = \frac{1}{V}\left\langle{\partial S_f \over \partial\mu_I}\right\rangle.
\end{equation}

The quark action for 2-colour QCD with one staggered quark is 
\begin{equation}
S_f = \sum_{sites}\left\{\bar{\chi}[D\!\!\!\!/\,(\mu) + m]\chi 
+ \frac{1}{2}\lambda[\chi^T\tau_2\chi + \bar{\chi}\tau_2\bar{\chi}^T]\right\}
\label{eqn:lagrangian}
\end{equation} 
where $D\!\!\!\!/\,(\mu)$ is the normal staggered covariant finite difference
operator with $\mu$ introduced by multiplying the links in the $+t$ direction
by $e^\mu$ and those in the $-t$ direction by $e^{-\mu}$. The superscript $T$
stands for transposition. The term proportional to $\lambda$ explicitly breaks
quark-number symmetry, and again we shall be interested in the limit 
$\lambda \rightarrow 0$. The symmetries of this action and the positivity of
the determinant and pfaffian have been discussed extensively in previous work
on dynamical quark simulations and will not be repeated here
\cite{hm,hklm,kts,kshm}.

The observables we measure include the chiral condensate,
\begin{equation}
\langle\bar{\chi}\chi\rangle=\langle\bar{\psi}\psi\rangle,
\end{equation}
the diquark condensate,
\begin{equation}
\langle\chi^T\tau_2\chi\rangle
\end{equation}
and the quark-number density
\begin{equation} 
j_0=\frac{1}{V}\left\langle{\partial S_f \over \partial\mu}\right\rangle.
\end{equation}

\section{Quenched simulations at finite chemical potentials}

\subsection{Quenched QCD at finite isospin chemical potential}

We have calculated quark propagators at a finite chemical potential $\mu_I$
for the third component, $I_3$, of isospin from a noisy source on an ensemble
of $12^3 \times 24$ and $16^4$ quenched gauge field configurations at
$\beta=5.7$ and on $8^4$ quenched gauge field configurations at $\beta=5.5$.
From these propagators we obtained stochastic estimators for the pion
condensate, the chiral condensate and the isospin density as functions of
$\mu_I$ and the $I_3$ breaking parameter $\lambda$. We chose $\lambda << m$,
since we are interested in the limit $\lambda \rightarrow 0$.

Let us first consider the $\beta=5.7$ calculations. We generated 100
$\beta=5.7$ $12^3 \times 24$ equilibrated configurations separated by 1000
sweeps consisting of 9 overrelaxation sweeps followed by 1 10-hit metropolis
sweep repeated 100 times, which appeared to generate relatively independent
configurations. On each of these configurations we obtained a stochastic
estimate of $i\langle\bar{\chi}\epsilon\tau_2\chi\rangle$,
$\langle\bar{\chi}\chi\rangle$ and $j_0^3$ using a single noisy source, for
$m=0.025$, $\lambda=0.0025,0.005,0.0075$ (and $\lambda=0$ for $\mu_I < \mu_c$) 
and $\mu_I=0.0,0.1,0.2,0.3,0.35,0.4,0.45,0.5,0.55,0.6,0.7,0.8,0.9,1.0,1.2,1.6,%
1.8,2.0,2.1$. The fact that we used the same set of configurations and noise 
vectors for each value of $\lambda$ and $\mu_I$ means that all our `data'
points are strongly correlated. We also measured the pion mass
at $\mu=\lambda=0$ using a wall source, on these configurations gauge fixed to
Coulomb gauge. We obtained $m_\pi = 0.441(1)$. Runs at $\lambda=0.0025$ were
performed on an $8^4$ lattice at selected $\mu_I$ values to check that finite
size effects were not too large.

From these $12^3 \times 24$ measurements we determined that the neighbourhood
of $\mu_I=0.45$ was of particular interest. We then increased our spatial
volume and performed simulations on a $16^4$ lattice, again at $\beta=5.7$ and
$m=0.025$, to allow us to run at smaller $\lambda$. Here we generated our
quenched configurations and performed our measurements of the condensates and
isospin density using the same hybrid molecular-dynamics code that we use for
simulations with dynamical quarks. This meant that we used an independent set
of configurations for each value of $\lambda$ and $\mu_I$. 100 independent
configurations separated by 10 molecular-dynamics time units were generated
for each $\lambda$ and $\mu_I$. (With no fermions in the updating, 10 time
units were adequate to decorrelate configurations. The time increment for
updating was $dt=0.1$ which is adequate without light fermions in the
updating.) We chose $\lambda=0.000625, 0.00125, 0.001875, 0.0025$, and used
$\mu_I=0.2,0.25,0.3,0.35,0.375,0.4,0.425,0.45,0.475,0.5,0.525,0.55,0.6,0.65,0.7$
to adequately cover the region of interest.

\begin{figure}[htb]
\vspace{-5pt}
\epsfxsize=4in
\centerline{\epsffile{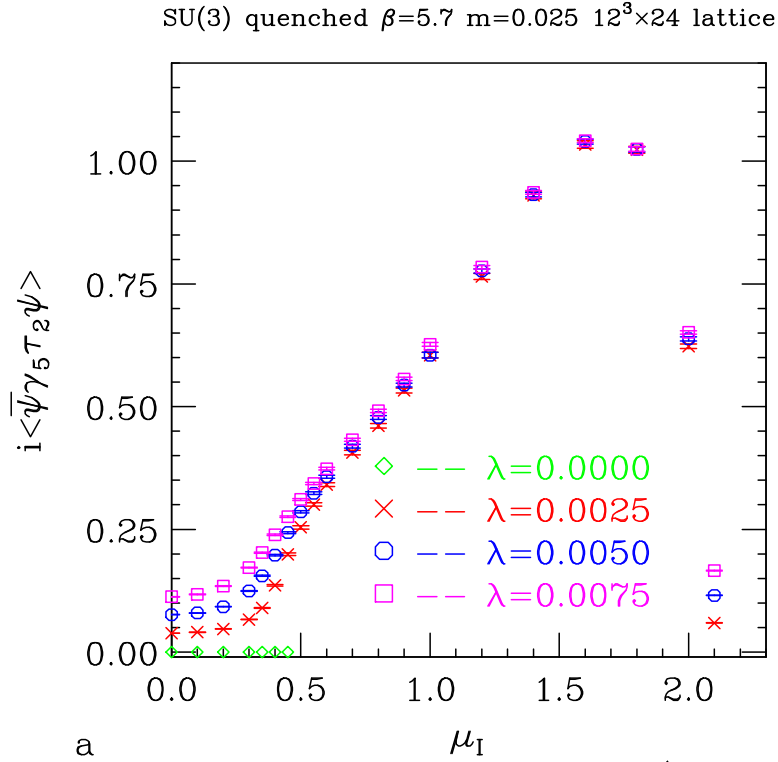}}
\centerline{\epsffile{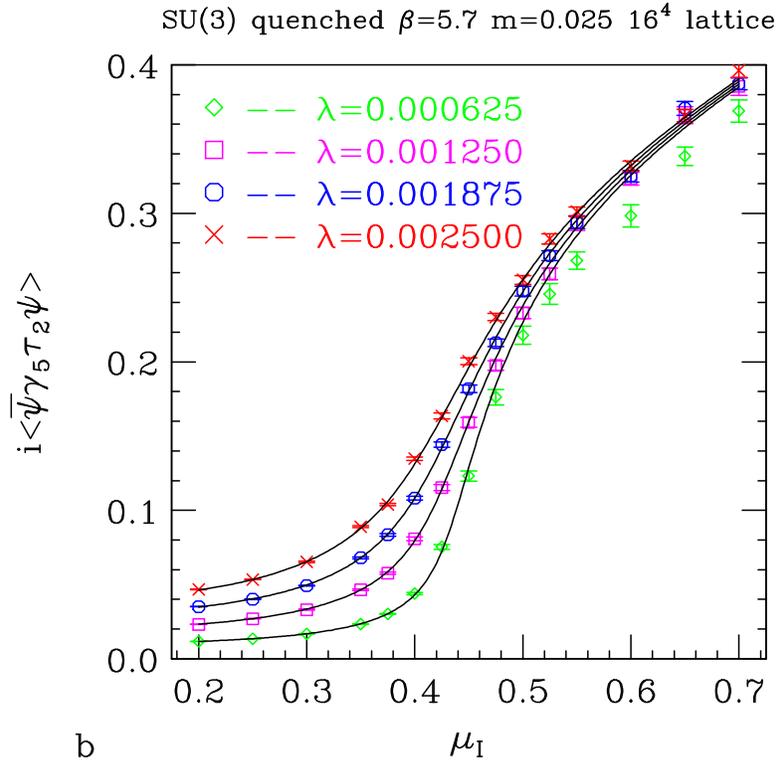}}
\caption
{a) Charged pion condensate, $i\langle\bar{\psi}\gamma_5\tau_2\psi\rangle =
i\langle\bar{\chi}\epsilon\tau_2\chi\rangle$, as a function of $\mu_I$, on a
$12^3 \times 24$ lattice with $\beta=5.7$ and quark mass $m=0.025$.
b) Charged pion condensate on a $16^4$ lattice. The solid lines superimposed on
our measurements are the scaling fits described in section~4.}
\label{fig:pion57}
\end{figure}
The charged pion condensates are shown in figure~\ref{fig:pion57}. What is
immediately clear is that the condensate is small for small $\mu_I$. At
$\mu_I=0$, the condensate vanishes as $\lambda \rightarrow 0$. Aloisio {\it et
al.}\cite{aadgg} have shown that this vanishing of the condensate extends to
small but finite $\mu_I$. We have performed various extrapolations to
$\lambda=0$ --- linear, linear plus cubic, linear plus quadratic -- all of
which are based on the predictions of partially conserved current analyses and
their implementations in effective (chiral) Lagrangian models. These
extrapolations suggest that this condensate vanishes out to at least $\mu_I
\approx 0.4$. (the $\lambda=0$ measurements are trivially zero and do not
enter into this discussion). More importantly, these extrapolation schemes,
and those based on power law scaling in $\lambda$ as would be expected for
$\mu_I$ at a critical value, indicate that it is highly probable that this
condensate remains finite as $\lambda \rightarrow 0$ in the infinite volume
limit for $\mu_I\gtrsim 0.5$. This indicates that there is a phase transition
for $\lambda=0$ at some $\mu_I=\mu_c$ with $0.4 \lesssim \mu_I \lesssim 0.5$,
below which the pion condensate is zero and above which there is a non-zero
charged pion condensate which breaks $I_3$ spontaneously with associated
Goldstone pions. Since there is no abrupt jump at $\mu_I=\mu_c$, this suggests
that the phase transition is second order. This condensate increases
monotonically up to $\mu_I \gtrsim 1.6$ after which it falls rather rapidly
approaching zero by $\mu_I \approx 2.0$. This fall is clearly a saturation
effect, a conclusion that becomes even more compelling when we look at $j_0^3$.

\begin{figure}[htb]                                  
\epsfxsize=6in
\centerline{\epsffile{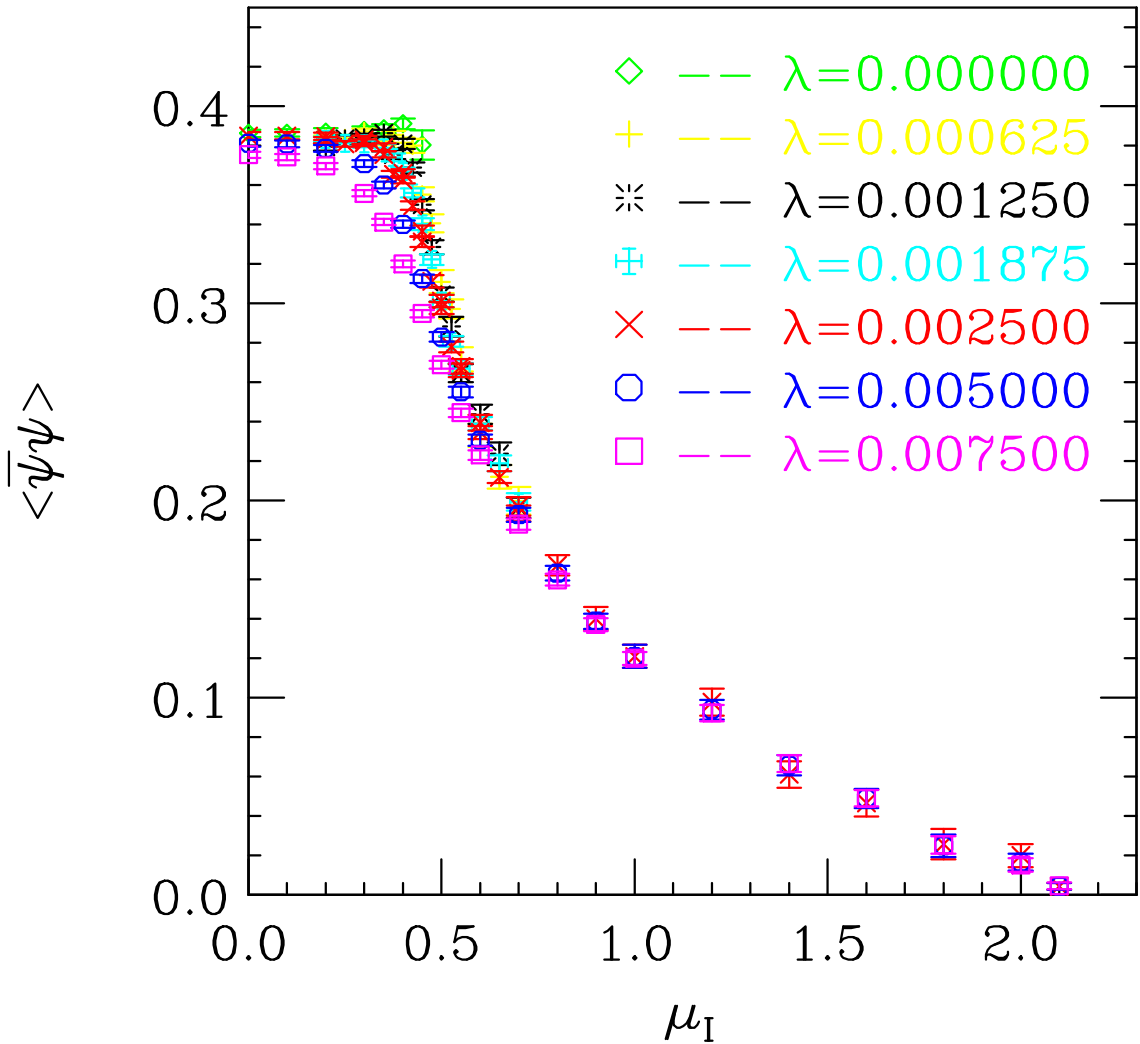}}
\caption{The chiral condensate,
$\langle\bar{\psi}\psi\rangle = \langle\bar{\chi}\chi\rangle$, 
at $\beta=5.7$ and $m=0.025$ as a function of $\mu_I$.}
\label{fig:pbp57}
\end{figure}
Figure~\ref{fig:pbp57} shows the chiral condensate as a function of $\mu_I$
for both the $12^3 \times 24$ and $16^4$ simulations at $\beta=5.7$, $m=0.025$.
We see that it remains approximately constant at its $\mu=0$ values for 
$\mu_I < \mu_c$, above which it commences its descent towards zero. What is
also clear when looking at the plots of chiral and pion condensates, is that
the expectation from tree level chiral perturbation theory that the condensate
merely rotates from the chiral symmetry breaking to the isospin breaking 
direction has at best a rather limited range of validity.

\begin{figure}[htb]
\epsfxsize=6in
\centerline{\epsffile{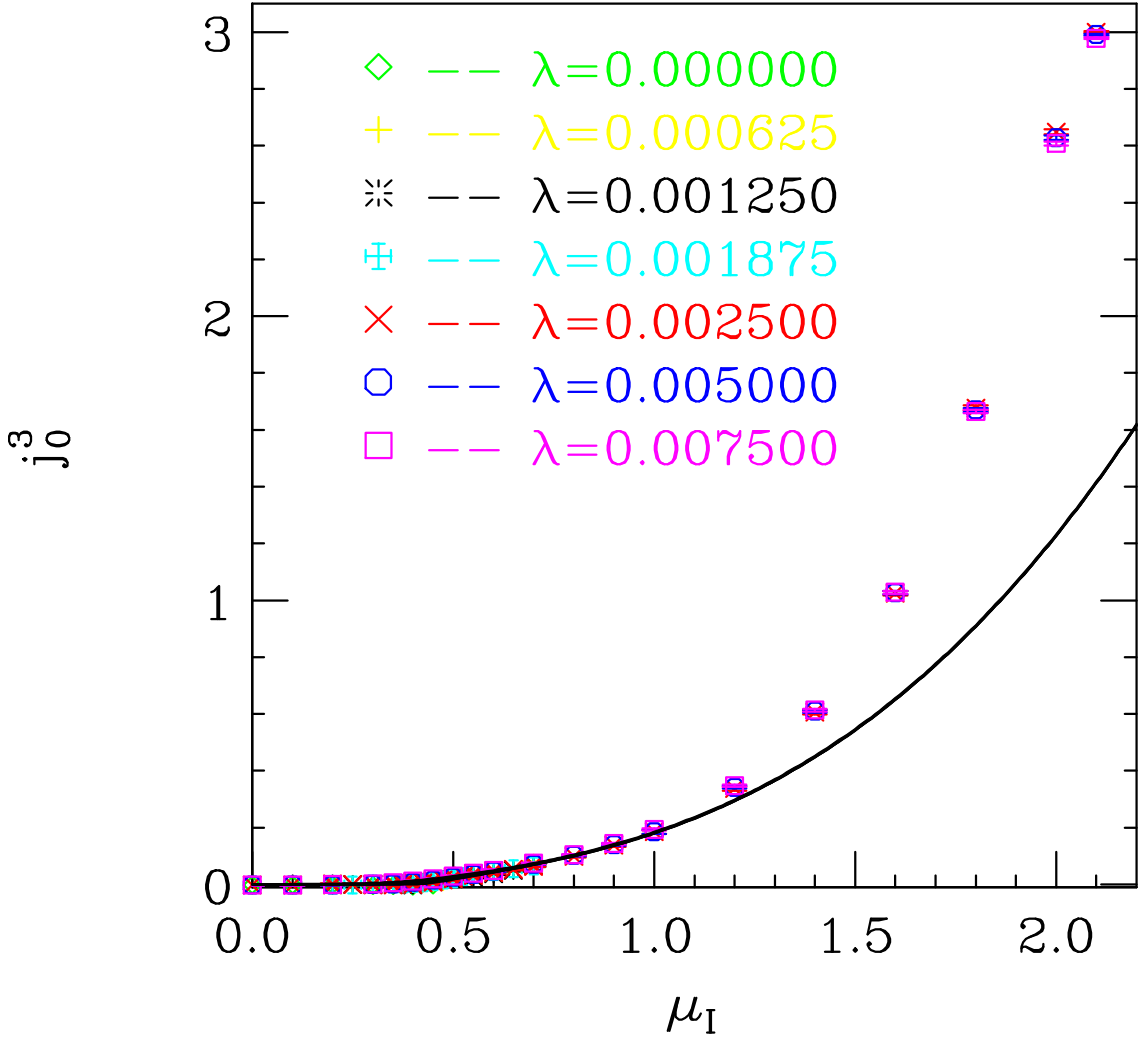}}
\caption{Isospin density at $\beta=5.7$ and $m=0.025$ as a function of $\mu_I$.
The curves are from the scaling fits to the pion condensate, and are described 
in section~4}%
\label{fig:j057}
\end{figure}
The isospin density is presented in figure~\ref{fig:j057}. The main features
are its slow rise from zero above $\mu_c$, followed by a more rapid rise at
larger $\mu_I$, finally flattening out at its saturation value of $3$ for
$\mu_I \gtrsim 2.0$. Note that the saturation value is $3$ because we have
chosen to normalize it to 8 continuum flavours, the number associated with
the action given in section 2. The value $3$ arises from $\frac{1}{2}$ for
each of 3 colours and 2 staggered `flavours' ($=8$ continuum flavours) on each
site, the maximum allowed by fermi statistics. (For contrast the condensates
have been normalized to 4 flavours for comparison with those reported in
finite temperature simulations.)

We now turn to the consideration of our simulations and measurements at 
stronger coupling, $\beta=5.5$, which were performed on an $8^4$ lattice. 
Again we chose $m=0.025$ in lattice units. Here we followed the same procedure
as for the $\beta=5.7$, $16^4$ simulations, using the same hybrid molecular-%
dynamics code as used for our dynamical simulations. We used
$\lambda=0.0025,0.005,0.0075$ (and $\lambda=0$ for $\mu_I < \mu_c$)in lattice
units. Our chosen $\mu_I$ (in lattice units) were $0.0,0.1,0.2,0.3,0.35,0.4,%
0.45,0.5,0.6,0.7,0.8,0.9,1.0,1.2,1.4,1.6,1.8,2.0,2.2$. We checked for finite
size effects, by repeating these simulations on a $12^4$ lattice for selected
$\mu_I$ values ($0.3$, $0.4$, $0.5$, $0.6$, $0.8$ and $1.0$), with
$\lambda=0.0025$ where such effects are expected to be largest, again
analyzing $100$ configurations for each $\mu_I$. For each $\mu_I$ the
difference between the $8^4$ and $12^4$ measurements was within 2 standard
deviations of zero.

\begin{figure}[htb]
\epsfxsize=6in
\centerline{\epsffile{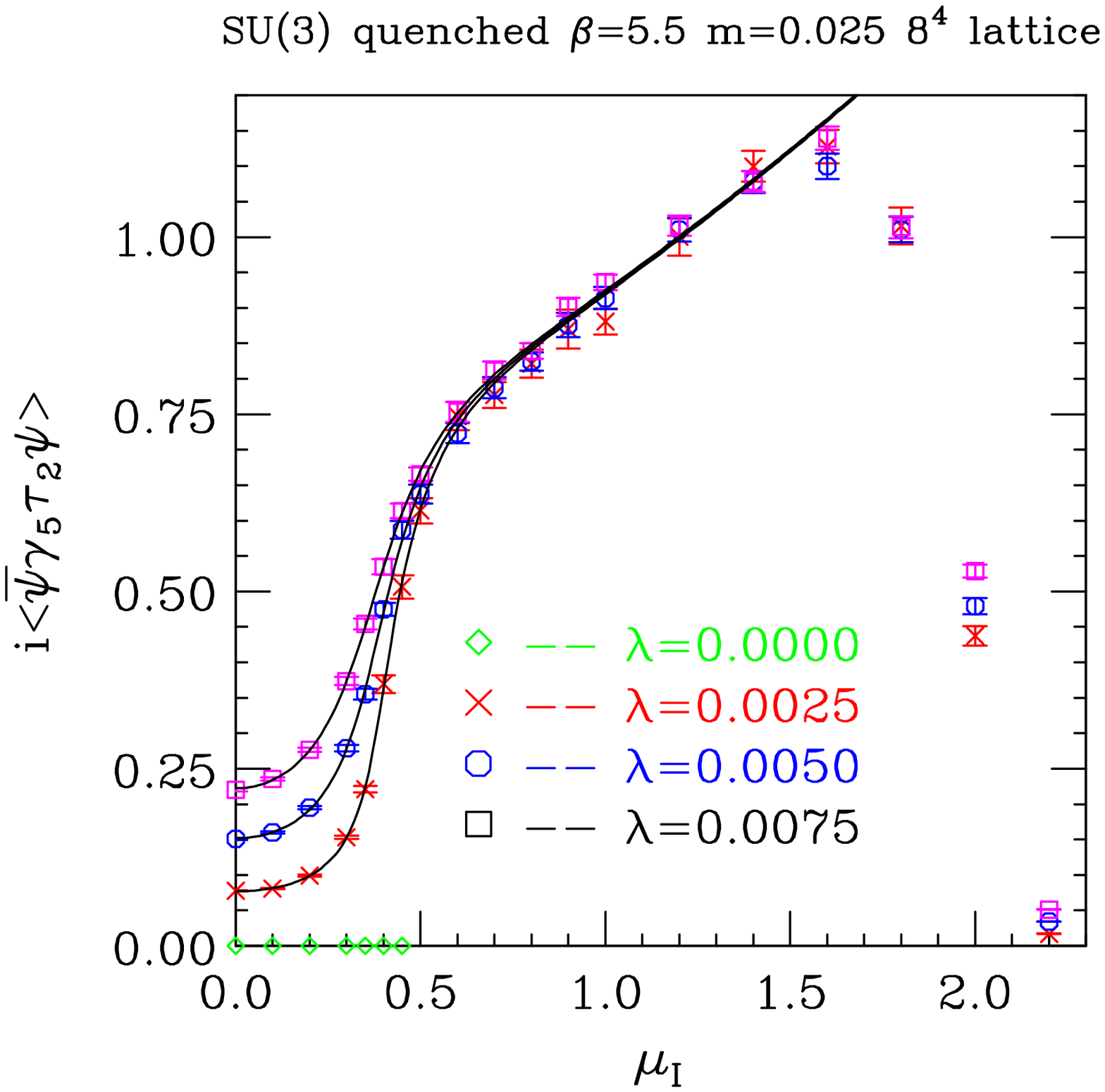}}
\caption{Charged pion condensate as a function of $\mu_I$ and $\lambda$ on
an $8^4$ lattice with $\beta=5.5$ and $m=0.025$. The solid lines are fits
described in section 4.}
\label{fig:pion55}
\end{figure}
The $I_3$-breaking charged pion condensate is plotted in figure~\ref{fig:pion55}
as a function of $\mu_I$, for the $\lambda$ values mentioned above.
Superficially, this graph appears similar to figure~\ref{fig:pion57}. Again,
we note that any reasonable extrapolation to $\lambda=0$ would suggest that, in
this limit, the condensate vanishes for $\mu_I \le 0.3$, while it clearly does
not for $\mu_I \ge 0.5$, which strongly suggests that there is a phase 
transition at $\mu_I=\mu_c$ with $\mu_c$ lying in the range 0.3--0.5. (We have
not performed such extrapolations since at larger $\mu_I$ values, statistical
fluctuations lead to incorrect ordering of the condensates for different
$\lambda$s, which would produce unphysical extrapolations.) However, what one
notices on closer inspection is that the transition is somewhat steeper than
that for $\beta=5.7$. We will return to this point in the next section.
Saturation again sets in for $\mu_I$ a little above $2.0$. 

\begin{figure}[htb]
\epsfxsize=6in
\centerline{\epsffile{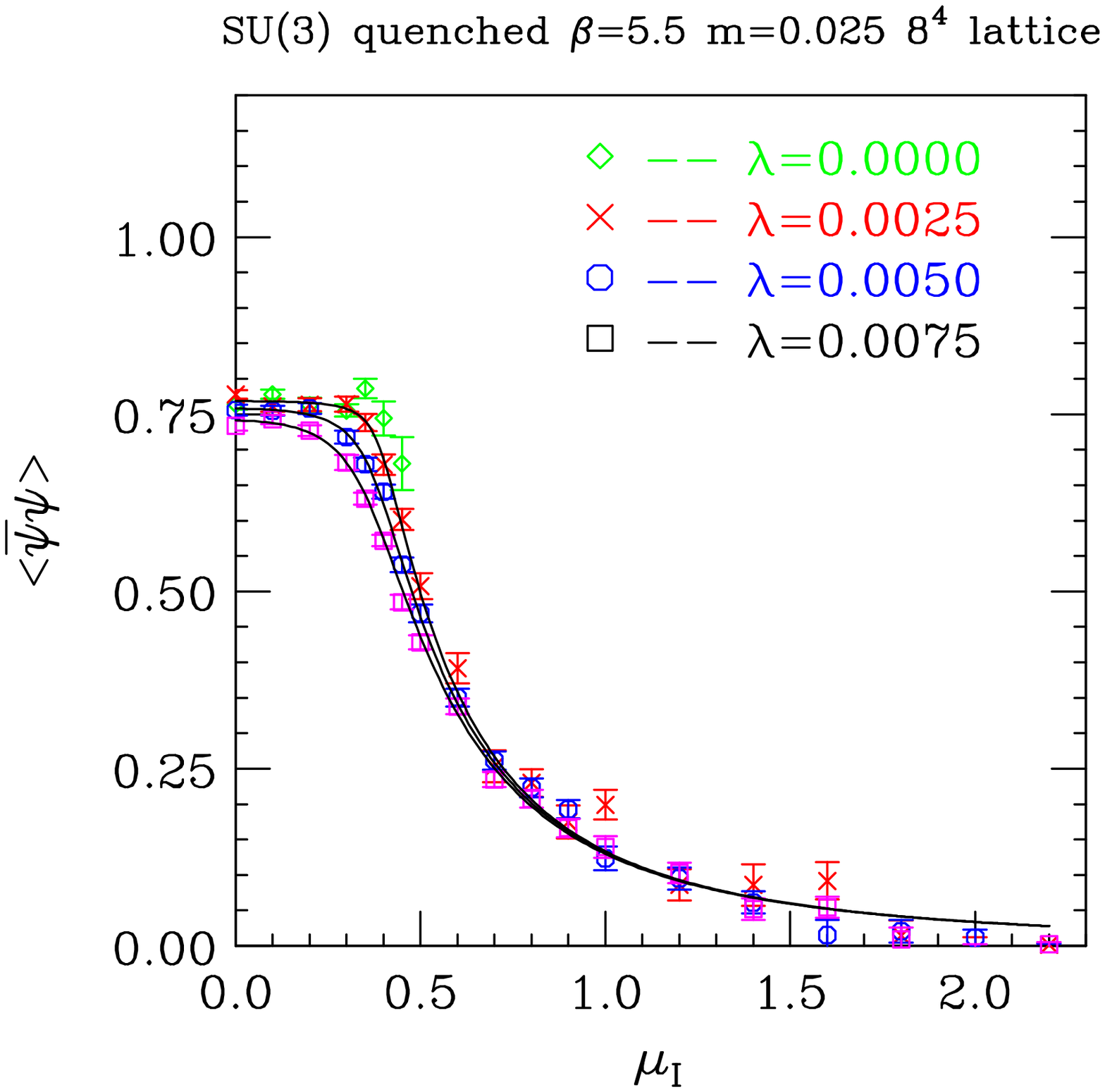}}
\caption{The chiral condensate as a function of $\mu_I$ for quenched QCD at
$\beta=5.5$, $m=0.025$. The solid lines are fits derived from the scaling
analyses of section~4.}%
\label{fig:pbp5.5}
\end{figure}

\begin{figure}[htb]
\epsfxsize=6in
\centerline{\epsffile{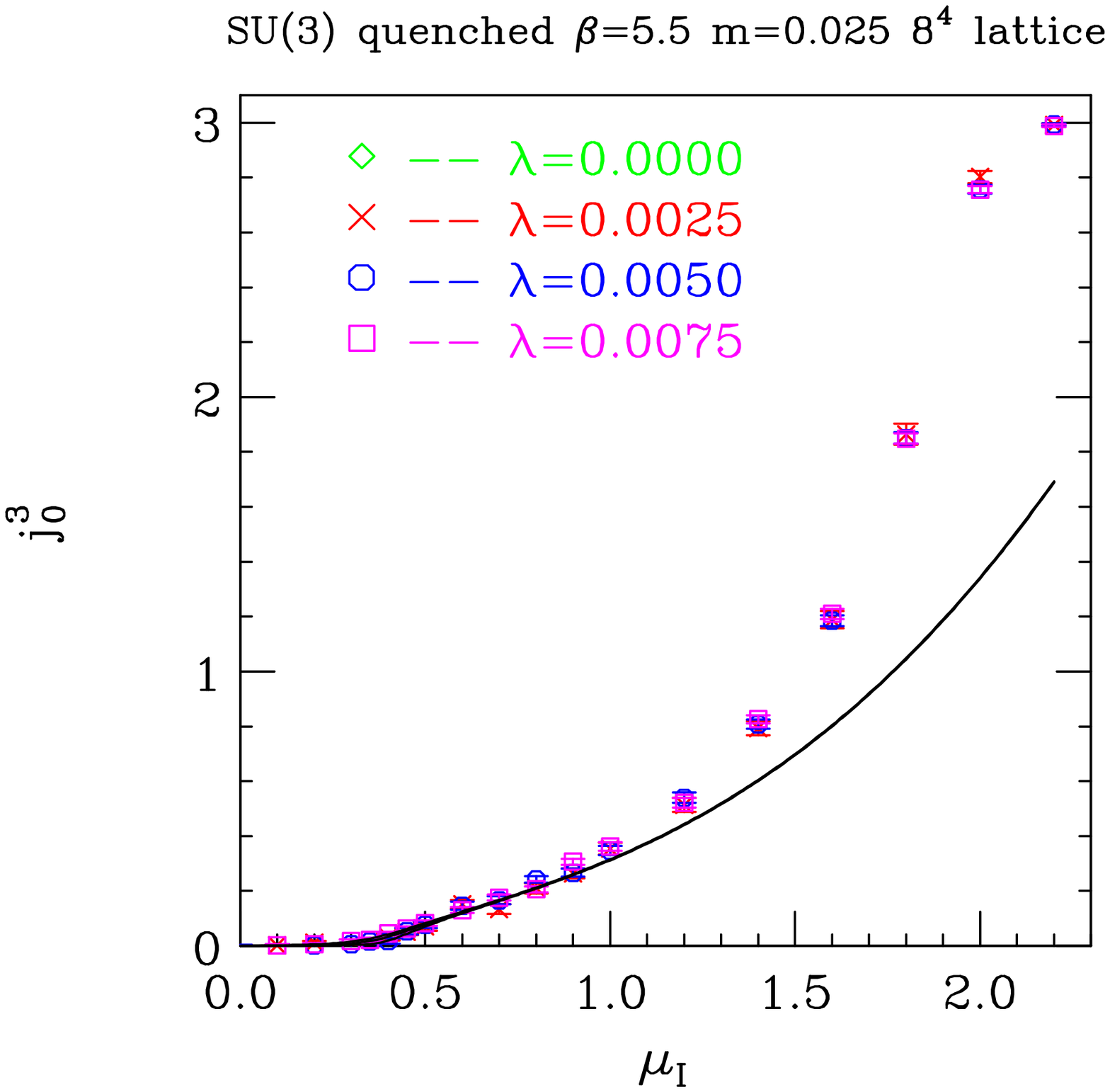}}
\caption{The isospin density as a function of $\mu_I$ for quenched QCD at
$\beta=5.5$, $m=0.025$. The solid lines are fits derived from the scaling
analyses of section~4.}%
\label{fig:j0_5.5}
\end{figure}

The chiral condensate shown in figure~\ref{fig:pbp5.5} behaves very similarly
to that at $\beta=5.7$. The main reason for presenting it here is to display
the scaling predictions of section~4. We note, however, that its value 
is $0.764(7)$ at $\lambda = \mu_I = 0$, so again it is at most over a limited
range of $\mu_I$ that the condensate simply rotates from the chiral towards
the isospin-breaking direction. However, the violations are less severe than
at $\beta=5.7$. The isospin density also behaves very similarly to that at
weaker couplings, rising steadily from zero as $\mu_I$ is increased past
$\mu_c$ and eventually saturating at $3$. This is plotted in
figure~\ref{fig:j0_5.5} again, primarily to show the scaling predictions
described in the next section.

\subsection{2-colour QCD at finite quark-number chemical potential.}

We calculated the quark propagators for fundamental staggered quarks at finite
chemical potential $\mu$ from a single noisy source for each of
a set of quenched SU(2) gauge field configurations at $\beta=2.0$. These
configurations were generated using the hybrid molecular-dynamics code of our
dynamical quark simulations. We generated 100 such independent gauge 
configurations spaced by 10 molecular-dynamics time units for each $\mu$ and
$\lambda$ value. Our simulations used $dt=0.1$ for the updating. We used
$\lambda = 0.0025, 0.005, 0.0075, 0.01$ (and zero for $\mu < \mu_c$), for
$\mu=0.0,0.1,0.15,0.175,0.2,0.21,0.225,0.25,0.3,0.4,0.5,0.6,0.7,0.8,0.9,0.95,%
1.0,1.1$, which adequately covered the range of interest. Since the $8^4$
`data' showed evidence of finite size effects, we repeated these simulations
on a $12^4$ lattice. From these noisy quark propagators we obtained stochastic
estimators for the diquark condensate, the chiral condensate and the
quark-number density, each normalized to 4 continuum flavours (1 staggered
quark field).

\begin{figure}[htb]
\epsfxsize=4in
\centerline{\epsffile{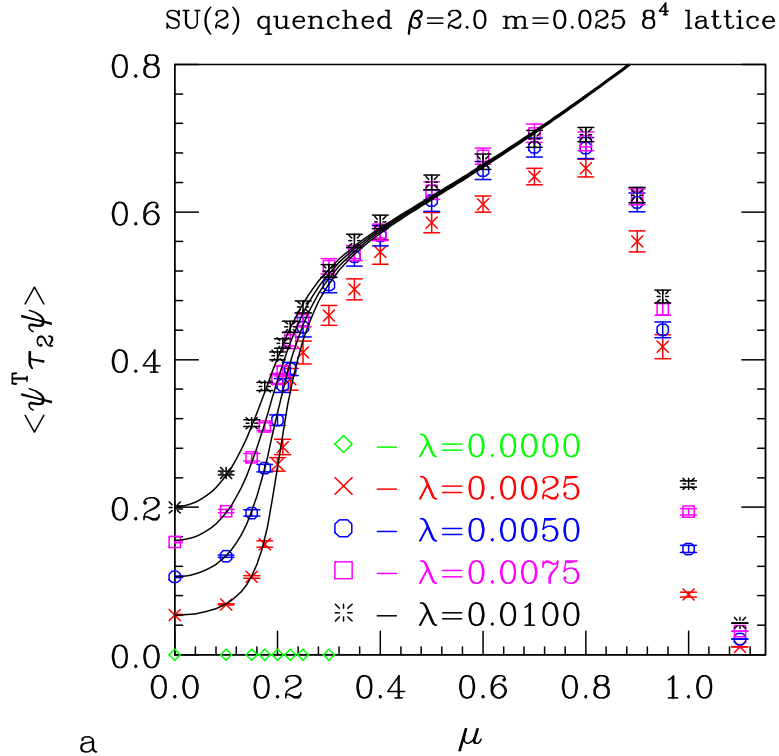}}
\centerline{\epsffile{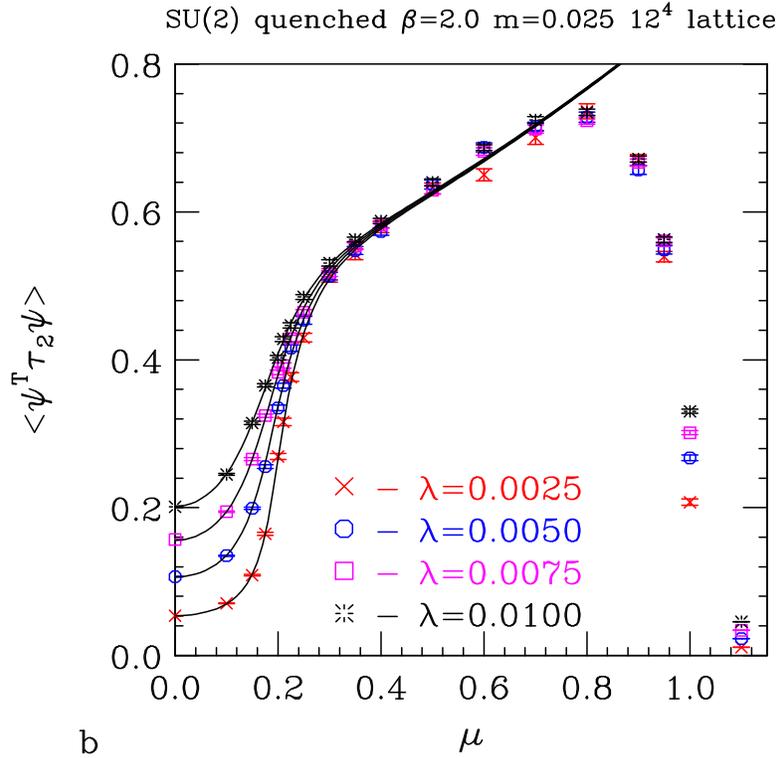}}
\caption{The diquark condensate for quenched 2-colour QCD as a function of 
$\mu$ and $\lambda$ on at $\beta=2.0$ a) on an $8^4$ lattice and b) on a
$12^4$ lattice. The curves are the scaling fits described in section~4.}%
\label{fig:diquark}
\end{figure}
Figure~\ref{fig:diquark} shows the diquark condensate as a function of $\mu$
for each of the considered $\lambda$ values. For $\mu \le 0.15$ it is clear
that any reasonable extrapolation to $\lambda=0$ from finite $\lambda$ would
yield estimates close to zero for the condensate in this limit. For 
$\mu \ge 0.25$, the extrapolated condensate clearly remains greater than zero
for $\lambda \rightarrow 0$. Thus we deduce that there is a phase transition
at $\mu = \mu_c$ from the normal state to one in which quark number is broken
spontaneously by a diquark condensate, for some $\mu_c$ between $0.15$ and
$0.25$. Such a symmetry breaking would be accompanied by a diquark Goldstone
boson. The reason we have not performed an extrapolation in $\lambda$ is clear
when one looks at the behaviour of the condensate at $\lambda=0.0025$ relative
to the higher lambdas. We note that this condensate increases monotonically
with $\mu$ up to $\mu \sim 0.7$ after which it falls due to saturation effects.
Qualitatively this curve is very similar to the corresponding curve with
dynamical quarks. It is also very similar to the curve for QCD at finite
isospin chemical potential at $\beta=5.5$ especially when one observes that we
should equate $\mu_I$ with $2\mu$. As in that case, the condensate increases
relatively rapidly as $\mu$ is increased beyond $\mu_c$. (This will be made
quantitative in the next section.) We note that the $\lambda=0.0025$ 
measurements on the $8^4$ lattice show signs of finite size effects for larger
$\mu$ values, which is why we repeated these simulations on a $12^4$ lattice
where these effects have been considerably reduced.

\begin{figure}[htb]
\epsfxsize=6in
\centerline{\epsffile{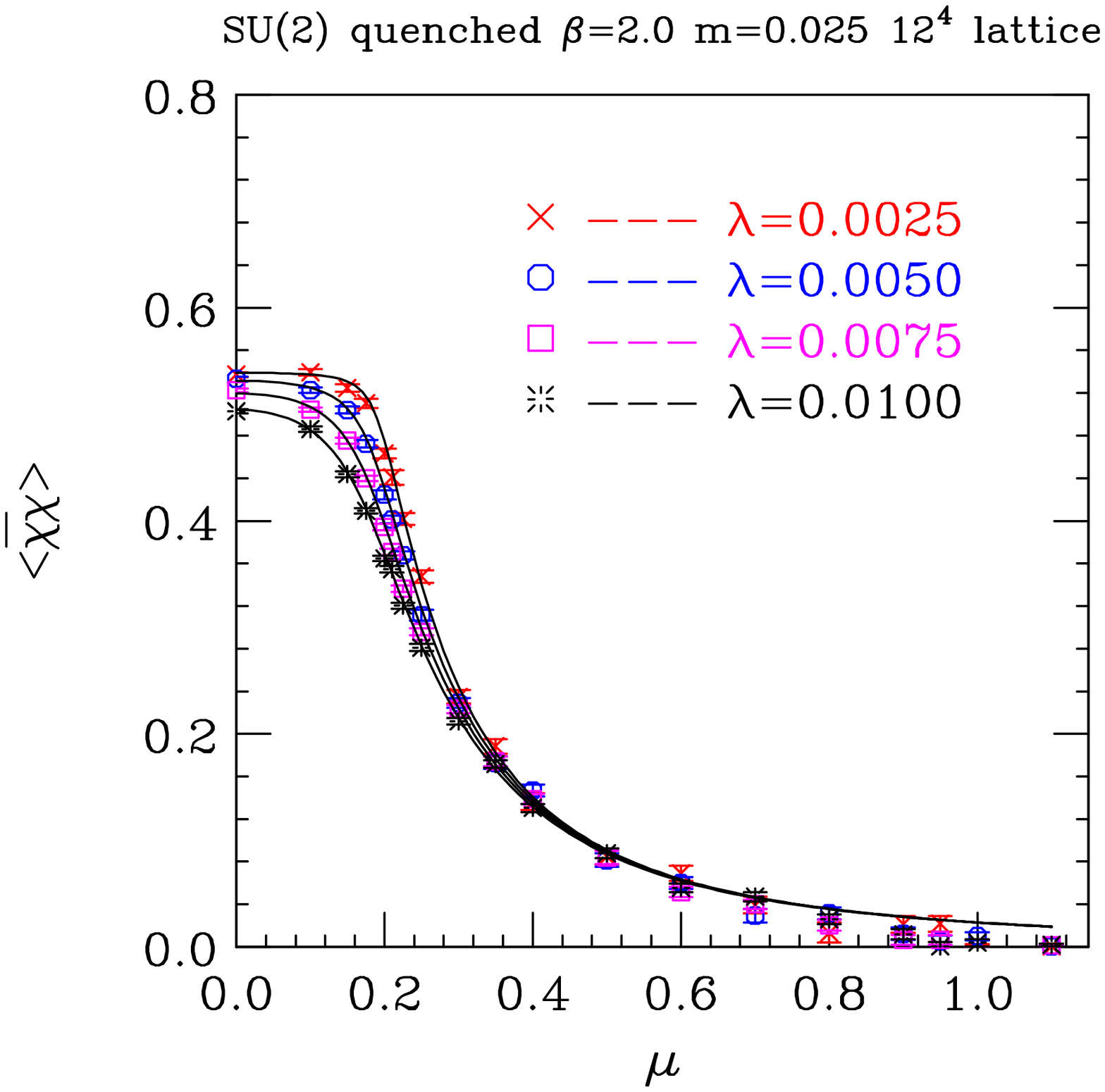}}
\caption{Chiral condensate as a function of $\mu$ and $\lambda$, for 2-colour
QCD at finite chemical potential for quark number on a $12^4$ lattice. The 
solid lines are the fits given in the next section.}%
\label{fig:pbp2}
\end{figure}
In figure~\ref{fig:pbp2} we show the chiral condensates as functions of $\mu$
for the 4 $\lambda$ values given above (and $\lambda=0$ for $\mu \le 0.2$). (We
present only the $12^4$ `data' since there is little difference between the 2
lattice sizes.) For $\mu < \mu_c$ this graph is consistent with the
expectation that the $\lambda \rightarrow 0$ limit should be $\mu$
independent. As $\mu$ is increased beyond $\mu_c$, the condensate falls. This
agrees with the expectation that the condensate is rotating from the chiral to
the diquark direction in this region. One notes, however, that this is not a
simple rotation since the magnitude of the condensate also increases with
$\mu$ for larger $\mu$ values until saturation effects appear. The general
appearance of this curve is similar to those with dynamical quarks and to
those for QCD at finite isospin density.

\begin{figure}[htb]
\epsfxsize=6in
\centerline{\epsffile{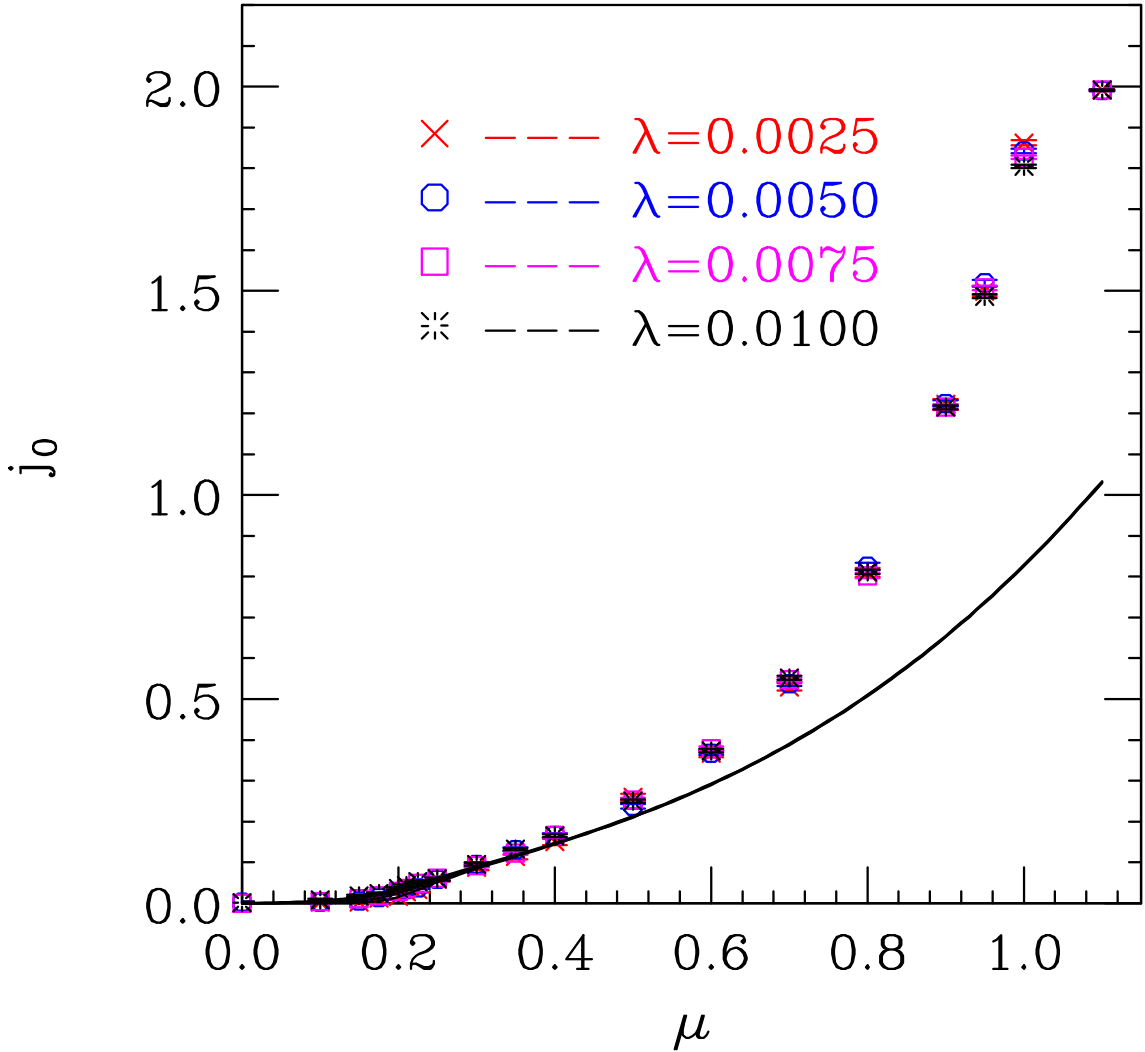}}
\caption{Quark-number density $j_0$ as a function of $\mu$ and $\lambda$ on}
\label{fig:j0}
\end{figure}
The quark-number density is shown in figure~\ref{fig:j0} as a function of $\mu$
for each of the considered $\lambda$ values mentioned above. (Again, because
the finite size effects are small, we present only the $12^4$ `data'). We see
that $j_0$ is consistent with zero for small $\mu$, and rises slowly from zero
above $\mu_c$. The rise becomes more rapid for $\mu \gtrsim 0.6$. The density
finally saturates at $2$ --- 2 colours of staggered fermions per site --- due to
fermi statistics, a finite lattice spacing effect. Note that there is little
if any $\lambda$ dependence. This behaviour is very similar to that of $j_0^3$
for our quenched QCD at finite $\mu_I$, except that there the saturation value
is $3$ rather than $2$ since this number merely counts colours. Again we note
the similarity of these quenched results to those obtained with dynamical
quarks.

The qualitative behaviour of all the `data' presented in this section is in
agreement with what is predicted from effective Lagrangians using the methods
of chiral perturbation theory. We have noted that the condensate does not
simply rotate from the chiral direction into the quark-number or isospin
violating direction which was the prediction of tree level chiral perturbation
theory \cite{ss,kstvz}, but rather rotates and increases in magnitude. Such
behaviour is seen when the chiral perturbation theory includes next-to-leading
order terms including 1-loop contributions \cite{stv}. Of course, the one
effect that effective Lagrangians cannot see is saturation, since they do not
describe the underlying fermions of the theory. However, as we have pointed
out, saturation is a finite lattice spacing effect, and is thus of limited
interest.

\section{Critical scaling and the equation of state.}

Analyses of effective Lagrangians using the methods of chiral perturbation
theory give predictions for the critical exponents and the scaling of the
order parameter(s) close to the critical point $\mu_c$, both for 2-colour QCD
at finite $\mu$ and for QCD at finite $\mu_I$. Clearly the tree level analyses
will give mean-field scaling with critical exponents $\beta_m=\frac{1}{2}$ and
$\delta=3$ \cite{ss,kstvz}. More recent analyses have extended these results
to next-to-leading order including 1-loop corrections for 2-colour QCD, and
find that the critical indices remain at their mean-field values \cite{stv}.
However, the tree-level result that the magnitude of the total condensate is
independent of $\mu$ and $\lambda$ no longer holds. One can thus speculate
that scaling with mean-field exponents holds to all orders. The effective
Lagrangian for QCD at finite $\mu_I$ is so similar as to lead to the
expectation that this theory also has mean field behaviour through
next-to-leading order and perhaps to all orders. Simulations of 2-colour QCD
at finite $\mu$ in the strong coupling limit \cite{aadgg} also indicate
mean-field scaling with a scaling function consistent with that of lowest
order tree-level effective Lagrangian predictions.

The two critical exponents relevant to the scaling of the charged pion
condensate --- the order parameter for the transition we are considering ---
are $\beta_m$ and $\delta$ defined as follows. At $m=0$,
\begin{equation}
i\langle\bar{\psi}\gamma_5\tau_2\psi\rangle 
\rightarrow const(\mu_I-\mu_c)^{\beta_m}
\end{equation}
as $\mu_I \rightarrow \mu_c +$ (and vanishes for $\mu_I \le \mu_c$). For
$\mu_I = \mu_c$, 
\begin{equation}
i\langle\bar{\psi}\gamma_5\tau_2\psi\rangle 
\rightarrow const \: m^\frac{1}{\delta}
\label{eqn:delta}
\end{equation}
as $m \rightarrow 0+$. The mean-field values of these 2 exponents are
$\beta_m = \frac{1}{2}$ and $\delta = 3$.
Analogous definitions hold for 2-colour QCD.

We shall be considering 3 different forms for the scaling function, i.e. for
the equation of state, each of which exhibits mean-field scaling. Indeed, 
sufficiently close to the critical point they give identical scaling functions
for the order parameter. Our later discussions will make it clear why, in
practical applications, it is important to have a more detailed knowledge of
the scaling function.

The first form for the equation of state is that suggested by the lowest order
tree-level analysis of the effective Lagrangians of \cite{ss,kstvz}, which are
of the non-linear sigma model form. It derives from the value of the parameter
$\alpha$ which minimizes the effective potential
\begin{equation}
{\cal E} = - a \: \mu^2 \: \sin^2(\alpha) - b \: m \: \cos(\alpha)
           - b \: \lambda \: \sin(\alpha)
\label{eqn:nls1}
\end{equation}
in terms of which
\begin{equation}
\langle\chi^T\tau_2\chi\rangle = b \: \sin(\alpha)
\label{eqn:nls2}
\end{equation}
\begin{equation}
\langle\bar{\chi}\chi\rangle = b \: \cos(\alpha)
\label{eqn:nls3}
\end{equation}
and
\begin{equation}
j_0 = 2 \: a \: \mu \: \sin^2(\alpha).
\label{eqn:nls4}
\end{equation}
$b$ is given in terms of $\mu_c$ and $a$, namely
\begin{equation}
b = \frac{2}{m} \: a \: \mu_c^2
\end{equation}
and the scaling forms of the observables for $\lambda=0$, expressed in terms of
the scaling variable $x = \mu/\mu_c$, are
\begin{equation}
\langle\chi^T\tau_2\chi\rangle = b \: \sqrt{1 - \frac{1}{x^4}}
\end{equation}
\begin{equation}                                      
\langle\bar{\chi}\chi\rangle = b \: \frac{1}{x^2} 
\end{equation}                                  
and
\begin{equation}                                
j_0 = 2 \: a \: \mu \: \left(1 - \frac{1}{x^4}\right)
\end{equation}                   
Note that the equivalent form for QCD at finite $\mu_I$ has an extra factor
of 2 in the equation for $j_0^3$ because of our normalization. Otherwise
the forms are identical with $\mu$ replaced with $\mu_I$. The main deficiency
with this form is that it requires that the magnitude of the condensate,
$\sqrt{\langle\chi^T\tau_2\chi\rangle^2 + \langle\bar{\chi}\chi\rangle^2}$
to be constant independent of $\mu$ and $\lambda$. This is not even true
in next-to-leading order chiral perturbation theory \cite{stv} and has at best
a limited range of validity in our simulations. 

To allow the norm of the condensate to increase with $\mu$ and possibly
$\lambda$, we consider a second form for the equation of state which is
derived from an effective Lagrangian of the linear sigma model type. This is
obtained by extracting the values of $R$ and $\alpha$ which minimize the
effective potential
\begin{equation}
{\cal E} = \frac{1}{4}\,R^4 - \frac{1}{2}\,a\,R^2 
         - \frac{1}{2}\,b\,\mu^2\,\sin^2(\alpha) 
         - c\,m\,R\,\cos(\alpha) - c\,\lambda\,R\,\sin(\alpha)
\label{eqn:ls1}
\end{equation}
in terms of which
\begin{equation}
\langle\chi^T\tau_2\chi\rangle = c \: R \: \sin(\alpha)
\label{eqn:ls2}
\end{equation}
\begin{equation}
\langle\bar{\chi}\chi\rangle = c \: R \: \cos(\alpha)
\label{eqn:ls3}
\end{equation}
and
\begin{equation}
j_0 = b \: \mu \: R^2 \sin^2(\alpha).
\label{eqn:ls4}
\end{equation}
$c$ is given in terms of $\mu_c$ by
\begin{equation}
c = { b \: \mu_c^2 \over m} \sqrt{a + b \: \mu_c^2}
\end{equation}
When $\lambda=0$ we obtain the scaling forms
\begin{equation}
\langle\chi^T\tau_2\chi\rangle = c \: \sqrt{a + b \: \mu_c^2 x^2 - 
                                          (a + b \: \mu_c^2)\frac{1}{x^4}}
\end{equation}
\begin{equation}
\langle\bar{\chi}\chi\rangle = c \: \sqrt{a + b\:\mu_c^2}\frac{1}{x^2} 
\end{equation}
and
\begin{equation}
j_0 = b \: \mu \: \left[a + b \: \mu_c^2 x^2 - (a + b \: \mu_c^2)\frac{1}{x^4}
                                                                       \right].
\end{equation}
Again $j_0^3$ has an extra factor of 2 for QCD at finite $\mu_I$.

Finally we consider the simplest mean-field equation of state \cite{textbook}
which considers only the scaling of the order parameter (diquark or
charged pion condensate) near the transition and does not even attempt to
address chiral symmetry breaking. This is given in terms of $\phi$ which
minimizes the effective potential
\begin{equation}
{\cal E} = \frac{1}{4}\phi^4 - \frac{1}{2}a(\mu-\mu_c)\phi^2 - b\,\lambda\,\phi.
\label{eqn:simple}
\end{equation}
In terms of $\phi$,
\begin{equation}
\langle\chi^T\tau_2\chi\rangle = b \:\phi
\end{equation}
and
\begin{equation} 
j_0 = a \: \phi^2.
\end{equation}
Once again $j_0^3$ has an extra factor of 2 for QCD at finite $\mu_I$. When
$\lambda=0$ we find the simple scaling relations for $\mu > \mu_c$
\begin{equation} 
\langle\chi^T\tau_2\chi\rangle = b \sqrt{a(\mu-\mu_c)} 
\end{equation}
and
\begin{equation}                                                                
j_0 = a^2 (\mu-\mu_c).
\end{equation}

We start by considering quenched QCD at finite $\mu_I$ at $\beta=5.5$. We first
fit our measurements of the pion condensate to the form suggested by the 
non-linear sigma-model effective Lagrangian (equations~\ref{eqn:nls1},%
\ref{eqn:nls2}). The quark mass is treated as a fitting parameter.
Here we find a good fit over the range $0 \le \mu_I \le 0.7$,
and all 3 $\lambda$ values with $\mu_c=0.422(2)$, $a=0.0618(8)$ and 
$m=0.0267(2)$ with a confidence level of $61\%$. The range limitation is due
to the fact that the norm of the condensate is only approximately constant
over a limited range of $\mu_I$. The departure of the mass $m$ from its
input value is presumably also a reflection of this. We then fit our pion
condensate `data' to the form suggested by the linear sigma-model effective
Lagrangian (equations~\ref{eqn:ls1},\ref{eqn:ls2}). With this form, our fits
constrain the mass parameter, $m$, to be so close to the input mass that
we choose to set $m=0.025$ in our fits. We then obtain an excellent fit over
to the pion condensate over the range $0 \le \mu_I \le 1.4$ with 
$\mu_c=0.419(2)$, $a=0.412(7)$, $b=0.242(4)$ and a confidence level of $92\%$.
We have superimposed these fits on the graphs of our `data' in 
figure~\ref{fig:pion55}. These fits give predictions for the chiral condensate
and isospin density (equations~\ref{eqn:ls3},\ref{eqn:ls4}) which we have
included in the plots of our measurements of these quantities in 
figures~\ref{fig:pbp5.5},\ref{fig:j0_5.5}. The predicted chiral condensate is
clearly in good agreement with our measurements. The predicted isospin density
is in good agreement with measurements up to $\mu_I \approx 1.0$. Note that
the isospin density has been predicted to be proportional to $\mu_I^3$ at
large $\mu_I$ \cite{ss} so that the predictions of equation~\ref{eqn:ls4} are
guaranteed to break down, even if we did not have the effects of saturation.

Let us now turn our attention to our simulations of 2-colour QCD at finite
$\mu$. First let us consider fits to our $8^4$ measurements. Here we again
note that our fits to the non-linear sigma model scaling form can only be
made over over a somewhat limited domain of $\mu$. This can be extended by
using the linear sigma model scaling forms, but only when we neglect the
$\lambda=0.0025$ measurements which show clear signs of finite size effects at
large $\mu$. Fitting the $\lambda=0.005,0.0075,0.01$ `data' we obtained a fit
over the range $0 \le \mu \le 0.7$ with $\mu_c=0.209(1)$, $a=0.405(9)$,
$b=0.71(2)$ and a $77\%$ confidence level. To reduce the finite size errors,
especially at $\lambda=0.0025$, we repeated these simulations on a $12^4$
lattice. Here our best fit, obtained using all 4 $\lambda$ values and fitting
over the range $0 \le \mu \le 0.4$, gave $\mu_c=0.2066(4)$, $a=0.405(9)$ and
$b=0.73(1)$ at a confidence level of $10\%$ ($\chi^2/dof=1.3$). Although we
have reduced our finite size effects by going to a larger lattice, we have
also reduced the statistical errors, making the remaining finite size effects
more important. We suspect that this is at least part of the reason that we
have obtained a worse fit on the larger lattice. These fits for the pion
condensate are shown in figure~\ref{fig:diquark}. The predictions for the
chiral condensate and quark-number density which these fits provide are
superimposed on the plots of measured values for the $12^4$ lattice in
figures~\ref{fig:pbp2},\ref{fig:j0}. It is clear that the prediction for the
chiral condensate provides good fits to the `data'. The predicted quark-number
density is also a good fit to the `data' for $\mu_I \le 0.4$.

Finally we turn to the consideration of the $\beta=5.7$ quenched QCD
simulations at finite $\mu_I$. After producing `data' over the range $0
\le\mu_I\le 2.1$ for $\lambda=0.0025,0.005,0.0075$ on a $12^3 \times 24$
lattice, we performed simulations on a $16^4$ lattice over the range $0.2
\le\mu_I\le 0.7$ at $\lambda=0.000625,0.00125,0.001875,0.0025$ to probe the
transition more closely. Not surprisingly, considering that the norm of the
condensate grows rapidly in the broken symmetry phase, fits to the non-linear
sigma model scaling proved fruitless. We fit our $16^4$ measurements
the linear sigma model equation of state, treating the quark mass as a fitting
parameter. We exclude the $\lambda=0.000625$ measurements from our fits since
they show clear finite size effects at larger $\mu_I$. Our `best' fit to the
$\lambda=0.00125,0.001875,0.0025$ measurements gives $\mu_c=0.4500(7)$,
$a=7.00(1)$, $b=0.266(7)$, $m=0.0212(4)$ with $\chi^2/dof=1.7$ and a confidence
level of only $0.3\%$. This indicates that while the linear sigma model
equation of state remains a good guide to the critical behaviour quenched QCD
at finite $\mu_I$ at $\beta=5.7$, the data is beginning to show small, but
non-negligible departures from this simple form. We have superimposed these
fits on the `data' of figure~\ref{fig:pion57} to indicate that they are a good
qualitative fit to the data. We note also that the predicted $\mu_c$ is close
to the measured pion mass $m_\pi=0.441(1)$.  Because $m \ne 0.025$ in these
fits, we do not expect the predictions for the chiral condensate to be good.
In fact, what we see is that the fits are suppressed relative to the data
roughly in the ratio of the fitted mass to the input mass ($0.025$). Next we
plot the values of $j_0^3$ predicted using equation~\ref{eqn:ls4} to compare
with measurements in figure~\ref{fig:j057}. We see that there is good
agreement out to $\mu_I=1.0$. Finally, we fit the pion condensate at
$\mu_I=0.45$ which is very close to $\mu_c$ to the scaling form of
equation~\ref{eqn:delta}. For this we use the measured values on the $16^4$
lattice at $\lambda=0.000625,0.00125,0.001875,0.0025$ and those from the 
$12^3 \times 24$ lattice at $\lambda=0.005,0.0075$. From this we obtain
$\delta=3.25(7)$ with a confidence level of $27\%$. Dropping the 2 largest
$\lambda$ values gives $\delta=2.88(16)$ with an $85\%$ confidence level. These
fits are in  good agreement with the mean field value $\delta=3$.

The effective Lagrangian approach seeks to give a uniform description of both
chiral symmetry breaking and isospin (or quark-number) breaking. Thus we should 
only expect physics to be insensitive to the details of the effective
Lagrangian/effective potential when the quark mass is small enough that the
order $m$ or equivalently order $m_\pi^2$ corrections are small. One measure
of this is how much the chiral condensate $\langle\bar{\psi}\psi\rangle$ at
the quark mass we use differs from its zero mass value. At $\beta=5.5$,
$\langle\bar{\psi}\psi\rangle(m=0.0125)=0.729(11)$,
$\langle\bar{\psi}\psi\rangle(m=0.025)=0.764(7)$ and
$\langle\bar{\psi}\psi\rangle(m=0.05)=0.838(5)$, so the difference is 
$\sim 10\%$ which should be small enough. At $\beta=5.7$,
$\langle\bar{\psi}\psi\rangle(m=0.00625)=0.288(3)$,
$\langle\bar{\psi}\psi\rangle(m=0.0125)=0.323(2)$,
$\langle\bar{\psi}\psi\rangle(m=0.025)=0.386(2)$ and
$\langle\bar{\psi}\psi\rangle(m=0.05)=0.494(2)$, so the difference is
$\sim 50\%$ and the detailed form of the effective potential might be expected
to be important. We believe that this is for this reason why the simple 
ansatz for the effective potential/equation of state works so well at
$\beta=5.5$ but shows limitations at $\beta=5.7$.

We end this section with a discussion of the behaviour of different variables
which describe the scaling in terms of $\mu(\mu_I)$. For convenience we express
them in terms of $x=\mu/\mu_c(\mu_I/\mu_c)$. The simplest is $x-1$ which is
the scaling variable of the simplest mean-field equation of state. The next,
$\frac{1}{4}(1-1/x^4)$, is that of the non-linear sigma model (or the linear
sigma model in the limit $a \rightarrow \infty$). The third,
$\frac{1}{6}(x^2-1/x^4)$ is that for the linear sigma model in the limit as 
$a \rightarrow 0$. This third form gives a reasonable description of the
scaling of $\beta=5.7$ `data' on the $12^3 \times 24$ lattice, in the ordered
($\mu_I \ge \mu_c$) phase. All 3 scaling variables behave identically as
$x \rightarrow 1$, however, a poor choice leads to a severely curtailed scaling
window. $x-1$ and $\frac{1}{4}(1-1/x^4)$ already differ by $\sim 10\%$ at
$x=1.04$. $x-1$ and $\frac{1}{6}(x^2-1/x^4)$ differ by $\sim 10\%$ at $x=1.08$.
Thus, for practical reasons, we should make a wise choice of scaling variables.

However, we have noted apparent mean-field scaling of our $\beta=5.7$ pion
condensate in the ordered phase both in terms of the simplest mean-field scaling
forms where the scaling variable is $x-1$, and in terms of the linear sigma
model form where the scaling variable is approximately $\frac{1}{6}(x^2-1/x^4)$,
out to $x \sim 1.8$. Only by examining the scaling very close to the transition
and including $\mu_I < \mu_c$ measurements in the fit, were we able to
determine that the linear sigma model form gave the better description of the
transition. The reason for this is that $\frac{1}{6}(x^2-1/x^4)$ has a point
of inflection at $x=\sqrt[6]{10}=1.467...$, at which it is again linear in
$x$. In the neighbourhood of this point of inflection
$\frac{1}{6}(x^2-1/x^4)\propto x-x_0$, where $x_0=(5/8)\sqrt[6]{10}=0.917...$,
so this defines another {\it larger} region where $\frac{1}{6}(x^2-1/x^4)$ is
linear in $x$, and the linear approximation has an intercept not too far from
the true critical point. It is this linearity, which gives us an (apparent)
extended scaling regime. Over the region where the scaling variable is
approximately linear in x, we will observe scaling with the correct exponent,
$\beta_m$. Of course, the coefficient of this apparent scaling form will
differ from the correct coefficient.

It is instructive to also look at the properties of $\frac{1}{4}(1-1/x^4)$ as
a function of $x$. It clearly does not have a point of inflection, however,
its square has a point of inflection at $x=\sqrt[4]{9/5}=1.158...$ about
which it too becomes linear. This linear approximation has its intercept at
$x_0=(9/10)\sqrt[4]{9/5}=1.042...$, which is again near to the critical point.
Since it is the square of the scaling variable that has this point of 
inflection, this leads to apparent scaling with a  critical exponent, $\beta_m$
of half the true value. In the case of a mean-field critical point this
pseudo-exponent will be $\frac{1}{4}$, which is $\beta_m$ for a tricritical
point. If we restrict ourselves to the ordered phase, it is possible to fit
the order parameter for QCD at finite $\mu_I$ at $\beta=5.5$ and that for
2-colour QCD at finite $\mu$ at $\beta=2.0$ to a tricritical scaling form ---
equation~\ref{eqn:simple} with $\frac{1}{4}\phi^4$ replaced by
$\frac{1}{6}\phi^6$. Hence one should be suspicious if an order parameter
scales with a critical exponent which is half of what one expects.

\section{Discussion and Conclusions}
 
Gauge theories with fermions at finite chemical potential for a conserved 
charge, which have positive fermion determinants, have sensible quenched
approximations. This contrasts with QCD at a finite chemical potential for
baryon/quark number, where the quenched limit is really the quenched limit
for QCD with equal numbers of quarks with quark-number $\pm 1$ and not that
for QCD with all quarks having quark-number $+1$.

We have studied the quenched versions of QCD at finite chemical potential 
($\mu_I$) for isospin, and of 2-colour QCD at finite chemical potential ($\mu$)
for quark number. We included an explicit symmetry breaking term with strength
$\lambda$ in each case. What we find is that for $\lambda \rightarrow 0$ this
symmetry --- $I_3$ for QCD at finite $\mu_I$, quark-number for 2-colour QCD ---
is unbroken for small $\mu(\mu_I)$. In each case there is a critical value for
$\mu_I$ or $\mu$ ($\mu_c$ say), above which a condensate forms which breaks
the relevant symmetry spontaneously. In our QCD at finite $\mu_I$ measurements
at $\beta=5.7$ on a $12^3 \times 24$ lattice, we also measured the pion mass
at $\mu_I=\lambda=0$, and found it to be close to our estimates for $\mu_c$ as
expected. In QCD at finite $\mu_I$ this is a charged pion condensate while for
2-colour QCD it is a diquark condensate. This is precisely the behaviour
expected and seen for the full (unquenched) theories. The transition appears
to be second order as was observed and expected in the full theories. For high
$\mu_I$($\mu$) saturation effects due to fermi statistics drive the
condensates to zero which is also observed for the full theories and is a
finite lattice spacing effect. Thus these quenched theories are useful to
study since they require far less computing time than the full theories ---
these studies were performed on 400MHz PC's supplemented with a few days
running on an IBM SP whereas the full theories require many months of IBM SP
running.

We studied the scaling of the order parameter,
$i\langle\bar{\psi}\gamma_5\tau_2\psi\rangle$ (the charged pion condensate),
for QCD at finite $\mu_I$, and $\langle\chi^T\tau_2\chi\rangle$ (the diquark
condensate) for 2-colour QCD at finite $\mu$ and saw evidence that the mean
field equation of state was the correct one in the neighbourhood of the
critical point. This required using scaling forms implied by effective
Lagrangian analyses, in order to use scaling variables which maximize the
scaling window and make scaling analyses practical. Because of this
sensitivity to the choice of scaling variables, these conclusions cannot be
considered conclusive. In the quenched theory, it might be feasible to extend
these simulations to much larger lattices and much smaller $\lambda$ values,
enabling one to probe sufficiently close to the critical point to be
insensitive to the choice of scaling variables. However, here one runs the
risk of being frustrated by the pathologies of quenching. For the dynamical
theories the expense (in computing resources) of simulating on large lattices
with very small $\lambda$ values is prohibitive. We have noted that even by
couplings as small as that at $\beta=5.7$, the simple scaling forms suggested
by effective Lagrangian analyses are barely adequate at the quark mass we use
($m=0.025$ in lattice units), and a better choice is desirable and probably
essential at weaker couplings. While it is easy enough to produce variant
forms of the effective potential and the equations of state they imply, one
needs a form which improves the fits, without introducing too many extra
parameters and preferably one motivated by physics or chiral perturbation
theory.

On examining the detailed properties of proposed scaling variables, we have
found that a certain class has a point of inflection, when expressed in terms
of the simplest scaling variable. This leads to an extended region of apparent
scaling which is a precursor of true scaling. In dynamical theories where one
cannot probe the true scaling window, this might be the best indicator we have
of the nature of critical scaling. However, as we have noted, certain other
proposed scaling variables which lack such a point of inflection, have squares
which {\it do} have such a point of inflection, leading to apparent scaling
with a critical index $\beta_m$ which is half of the true value, potentially
leading one to erroneous conclusions.

Once the fit to the pion or diquark condensate provides the parameters for the
appropriate effective potential, this provides a prediction for the isospin
or quark-number density. These predictions are in good agreement with our 
direct measurements of these `charge' densities within the scaling window. In
addition, the effective potential provides a prediction of the chiral
condensate. For quenched QCD at finite $\mu_I$ at $\beta=5.5$ and quenched
2-colour QCD at finite $\mu$ at $\beta=2.0$, where the fits require a quark
mass $m$ in good agreement with the input value ($m=0.025$), the predicted
chiral condensates are in good agreement with the measured values. In 
quenched QCD at finite $\mu_I$ at $\beta=5.7$, where the fits require $m$ to
be appreciably lower than the input value, the predictions for the corresponding
chiral condensate are similarly lower. This we tentatively attribute to the
fact that the higher order terms in a chiral expansion (expansion in powers of
$m$) are large at $\beta=5.7$, $m=0.025$. Where the predictions from our
proposed equation of state are in good agreement with measurements, they
support our claim that there is a critical point at $\mu(\mu_I)=\mu_c$, 
$\lambda=0$, with mean field exponents.
 
We have just completed further simulations of 2-colour QCD at finite
quark-number chemical potential with dynamical quarks which examine the
critical scaling at the transition from the normal phase to the
diquark-condensed phase in which we also see evidence for mean-field scaling
\cite{kts2}. In addition, we are performing simulations of QCD at finite
chemical potential for isospin($I_3$) \cite{lattice2001,qcdi} and see evidence
for mean field scaling \cite{qcdi}.

We are extending the work of this paper to include a strange quark with its
own chemical potential. This work will study the competition between charged
pion and kaon condensation as the 2 chemical potentials are varied 
independently. Chiral perturbation theory predicts that the phases with pion
and kaon condensates are separated by a first order phase transition. Here
the quenched approximation will make it easy to study the phase structure of
the plane defined by the 2 chemical potentials.

\section*{Acknowledgements}

DKS was supported by DOE contract W-31-109-ENG-38. JBK was supported in part
by an NSF grant NSF PHY-0102409. IBM SP access was provided at NERSC. DKS
would like to thank the Special Research Centre for the Subatomic Structure of
Matter of the University of Adelaide where this work was completed. He would
also like to thank C.~K.~Zachos and G.~T.~Bodwin for helpful discussions. JBK
wishes to thank D.~Toublan for many useful discussions.


\begin{thebibliography}{9999}
\bibitem{ss}
D.~T.~Son and M.~A.~Stephanov, Phys. Rev. Lett. 86, 592 (2001);
D.~T.~Son and M.~A.~Stephanov, Phys. Atom. Nucl. 64, 834 (2001), 
Yad. Fiz. 64, 899 (2001). 
\bibitem{lattice2001}
J.~B.~Kogut and D.~K.~Sinclair, Nucl. Phys.(Proc. Suppl.) 106, 444, (2002).
\bibitem{qcdi}
J.~B.~Kogut and D.~K.~Sinclair, e-print hep-lat/0202028 (2002).
\bibitem{hm}
S.~E.~Morrison and S.~J.~Hands, in "Strong and Electroweak Matter '98'," 
edited by J. Ambj{\o}rn et al., p. 364.
\bibitem{hklm}
S.~Hands, J.~B.~Kogut,~M.-P.~Lombardo, and S.~E.~Morrison, Nucl. Phys. B558, 
327 (1999).
\bibitem{kts}
J.~B.~Kogut, D.~Toublan, and D.~K.~Sinclair, Phys. Lett. B514, 77 (2001) 
\bibitem{kshm}
J.~B.~Kogut, D.~K.~Sinclair, S.~J.~Hands, S.~E.~Morrison,
Phys. Rev. D64, 094505 (2001).
\bibitem{aadgg}
R.~Aloisio, V.~Azcoiti, G.~Di~Carlo, A.~Galante, and A.~F.~Grillo, 
Phys. Lett. B 493, 189 (2000);
R.~Aloisio, V.~Azcoiti, G.~Di~Carlo, A.~Galante, and A.~F.~Grillo,
Nucl. Phys. B606, 322 (2001).
\bibitem{mnn}
S.~Muroya, A.~Nakamura, C.~Nonaka, e-print nucl-th/0111082 (2001). 
\bibitem{kstvz}
J.~B.~Kogut, M.~A.~Stephanov and D.~Toublan, Phys. Lett. B464, 183 (1999).
J.~B.~Kogut, M.~A.~Stephanov, D.~Toublan, J.~J.~M.~Verbaarschot and 
A.~Zhitnitsky, Nucl. Phys. B582, 477 (2000).
\bibitem{stv}
K.~Splittorff, D.~Toublan and J.~J.~M.~Verbaarschot, e-print hep-ph/0108040
(2001).
\bibitem{b} 
B.~C.~Barrois, Nucl. Phys. B129, 390 (1977)
\bibitem{bl}
D.~Bailin and A.~Love, Phys. Rept. 107, 325 (1984).
\bibitem{arw}
M.~Alford, K.~Rajagopal and F.~Wilczek, Phys. Lett. B222, 247 (1998);
M.~Alford, K.~Rajagopal and F.~Wilczek, Nucl. Phys. B537, 443 (1999);
T.~Sch\"{a}fer and F.~Wilczek, Phys. Rev. D60, 074014 (1999);
T.~Sch\"{a}fer and F.~Wilczek, Phys. Rev. D60, 114033 (1999).
\bibitem{rssv}
R.~Rapp, T.~Schafer, E.~V.~Shuryak and M. Velkovsky, Phys. Rev. Lett. 81, 53
(1998)
\bibitem{as}
Good recent reviews include: M.~Alford, Nucl. Phys. B(Proc. Suppl.) 73, 161 
(1999); E.~Shuryak, Nucl. Phys. B(Proc. Suppl.) 83-84, 103 (2000);
K.~Rajagopal and F.~Wilczek, e-print hep-ph/0011333 (2000).
\bibitem{barbour}
I.~Barbour, Nucl. Phys. B(Proc. Suppl.)26, 22 (1992).
\bibitem{bbdkmsw}
I.~Barbour, N.~E.~Behilil, E.~Dagotto, F.~Karsch, A.~Moreo, M.~Stone, and 
H.~W.~Wyld, Nucl. Phys. B 275 [FS17], 296 (1986).
\bibitem{dk}
C. T. H. Davies and E. G. Klepfish, Phys. Lett. B 256, 68 (1991), 
and references contained therein.
\bibitem{stephanov}
M.A. Stephanov, Phys. Rev. Lett. 76, 4472 (1996). 
\bibitem{kls} 
J.~B.~Kogut, M.-P.~Lombardo and D.~K.~Sinclair, Phys. Rev. D51,
1282 (1995); Phys. Rev. D54, 2303 (1996).  
\bibitem{textbook}
See for example: C.~Itzykson and J-M~Drouffe, ``Statistical Field Theory'',
Vol.~1, Chapter~3, Cambridge University Press (1989).
\bibitem{sbg}
S.~R.~Sharpe, Phys. Rev. D41, 3233 (1990);
S.~R.~Sharpe, Phys. Rev. D46, 3146 (1992);
C.~Bernard and M.~Golterman, Nucl. Phys. B(Proc. Suppl.)26, 360 (1992);
C.~W.~Bernard and M.~F.~L.~Golterman Phys. Rev. D46, 853 (1992).
\bibitem{xqcd}
J.~B.~Kogut and D.~K.~Sinclair, Phys. Lett. B492, 228 (2000); 
J.~B.~Kogut and D.~K.~Sinclair, Phys. Rev. D64, 034508 (2001). 
\bibitem{dl-ls} 
I.~D.~Lawrie and S.~Sarbach, in {\it Phase Transitions and
Critical Phenomena, Volume 9} (Academic Press, London, 1984).
\bibitem{kts2}
J.~B.~Kogut, D.~Toublan, and D.~K.~Sinclair (in preparation).

\end{thebibliography}
\end{document}